\documentclass[twocolumn,showpacs,fleqn,nobibnotes]{revtex4}

\usepackage{amsmath}
\usepackage{graphicx}
\usepackage{float}
\usepackage{subfigure}

\def\lsim{\raise0.3ex\hbox{$<$\kern-0.75em\raise-1.1ex\hbox{$\sim$}}}
\def\gsim{\raise0.3ex\hbox{$>$\kern-0.75em\raise-1.1ex\hbox{$\sim$}}}

\begin{document}

\title{On theoretical uncertainty of color dipole phenomenology in the $J/\psi$ and $\Upsilon $ photoproduction in $pA$ and $AA$ collisions at the  CERN Large Hadron Collider}
\author{G. Sampaio dos Santos and M.V.T. Machado}

\affiliation{High Energy Physics Phenomenology Group, GFPAE  IF-UFRGS \\
Caixa Postal 15051, CEP 91501-970, Porto Alegre, RS, Brazil}

\begin{abstract}
We investigate the theoretical uncertainty on the predictions for the photoproduction of $J/\psi$ and $\Upsilon $ states in the proton-nucleus  and nucleus-nucleus collisions at the LHC within the color dipole formalism. Predictions for the rapidity distributions are presented and the dependence on the meson wavefunction, heavy quark mass as well as the models for the dipole cross section are analyzed. We compare directly the theoretical results to the recent data from ALICE collaboration on $J/\psi$ production in pPb collisions at a centre-of-mass energy per nucleon pair of  5.02 TeV and in PbPb  collisions at a centre-of-mass energy per nucleon pair of 2.76 TeV. Predictions are also performed for $\Upsilon$ state in PbPb  and pPb collisions at the LHC energies, including the coherent and incoherent contributions.

\end{abstract}

\maketitle

\section{Introduction} 

The recent measurements of the exclusive quarkonium photoproduction by ALICE \cite{ALICE1,ALICE2,ALICE3} and LHCb \cite{LHCb1,LHCb2} collaborations at the LHC have imposed serious constraints to the theoretical models for vector meson (V) production (see a review in Ref. \cite{Tapia}). The forward and central rapidities regions are mapping the dynamics at very small Bjorken-$x$ values, $x \simeq (M_V/\sqrt{s})e^{\pm y}$ (here, $M_V$  and $y$ are the meson mass and rapidity, respectively). The centre-of-mass energy of hadron-hadron collision is denoted by $\sqrt{s}$. For instance, for the $J/\psi$ this means values of $x\simeq 10^{-4}\,(10^{-3})$ at central rapidity in proton-proton collisions at $\sqrt{s}=7$ TeV (in lead-lead collisions at centre-of-mass energy per nucleon pair $\sqrt{s_{NN}}=2.76$ TeV). Accordingly, the theoretical uncertainty in these cases is very large as the underlying models basically depend on the gluons distribution squared, which is uncertain at the small-$x$ regime. The situation is more complicated in the nuclear case, where there is the additional problem of determining the nuclear gluon distribution function. A very promising approach is given by the color dipole framework \cite{nik,Nemchik:1996pp}, where the inclusive and exclusive processes can be treated simultaneously. Moreover, the dipole framework allows to include in a simple way the corrections associated to the parton saturation phenomenon \cite{hdqcd} and to introduce information on dynamics beyond the leading logarithmic QCD approach.

As a summary on our recent works in the subject, in Ref. \cite{GGM1} the photoproduction of $J/\psi$ and $\psi (2S)$ states were computed in color dipole formalism and the results are in good agreement with LHCb data \cite{LHCb1,LHCb2} on $pp$ collisions at 7 TeV. On the other hand, in Ref. \cite{GGM2} the same states were analyzed in the lead-lead collisions at 2.76 TeV, including the coherent and incoherent contributions. The predictions are somewhat consistent with the ALICE data \cite{ALICE1,ALICE2}. It was verified that the theoretical uncertainty is very large compared to the proton case, mostly at central rapidities. The quarkonium production, i.e. the $J/\psi$ and $\Upsilon$ states, in proton-lead collisions were investigated in Ref.\cite{Glauber1}. There, it was included predictions for the rapidity distribution in backward and forward regions in the process $Pb+p\rightarrow Pb+p+V$. In the proton-lead case it is possible to investigate  at the same
time the dynamics on the photon-proton cross section and on the
photon-nucleus cross section. The dominant contribution comes from the photon-proton interaction as the photon flux due to the nucleus is higher compared to that due to the proton. It was shown in \cite{Glauber1} that the photon-nucleus contribution is relevant at large rapidities and increasingly important for the $\Upsilon$ states. The present work is an extension of Ref. \cite{Glauber1}, where  the focus is on the investigation of the theoretical uncertainties associated to the models for the dipole cross section and meson wavefunction. Furthermore, we provide predictions for the exclusive $\Upsilon$ production in coherent and incoherent lead-lead collisions at the LHC including their theoretical uncertainties. We note that in the present work we consider the process $p+Pb\rightarrow p+Pb+V$ as recently been measured by ALICE collaboration \cite{ALICE3} in 5.02 TeV energy. 

Here, we will consider the color dipole framework \cite{nik,Nemchik:1996pp}, which represents a QCD motivated formalism that has been successfully applied to describe a large variety of gluon mediated scattering cross sections at small-$x$ regime. Such a formalism provides a transparent and intuitive picture of scattering processes, having an universal and efficient character as the dipole models are able to simultaneously describe the inclusive $ep$ data (structure functions $F_{2}$, $F_{L}$ and heavy quark production), the diffractive DIS data (diffractive structure functions, $F_2^D$) and the bulk of measurements for exclusive production in $ep$ collisions (vector meson production, deeply virtual Compton scattering - DVCS).  The formalism also provides an easy way to introduce the corrections of higher  twist contribution to the  cross sections and the effects of multiple scattering (connected to the physics of parton saturation). The last feature is directly related to the physics of nuclear shadowing when nuclear targets at high energies are considered. The main idea behind the color dipole framework is the factorization of the high energy scattering amplitude into a initial and final state wavefunction of the projectile and outgoing state and a universal scattering amplitude of a multi-parton Fock state off a target (protons or nuclei).In practical applications, the considered Fock state is the quark-antiquark pair (the color dipole).

The structure of the paper is as follows. In next section we summarize the main expressions from the theoretical approach considered in present study. In Sec. III, we present the phenomenological inputs used  and discuss the models for the meson wavefunction and dipole cross sections. In Sec. IV, the main theoretical results are shown, comparing the predictions to experimental data when possible. In last section we summarize the main results and address the conclusions on the study performed.

\section{Theoretical framework} 

 In proton-nucleus interactions at large impact parameter ($b > R_p + R_A$, with $R_p$ and $R_A$ being the proton and nucleus radius, respectively) and at ultra-relativistic energies it is expected that the electromagnetic interaction will  be dominant. In this case, the exclusive meson photoproduction in hadron-target collisions can be factorized in terms of the equivalent flux of photons of the hadron projectile and photon-target production cross section \cite{upcs}. The photon energy spectrum for protons and nuclei, $dN_{\gamma}^p/d\omega$ and $dN_{\gamma}^A/d\omega$, which depend on the photon energy $\omega$,  are well understood (see details for the photon flux on protons or nuclei in Ref. \cite{upcs}). For proton-lead collisions at the LHC energies,  where the meson rapidity is positive in the proton beam direction, the quarkonium rapidity distribution reads as \cite{upcs}:
\begin{eqnarray}
\frac{d\sigma }{dy} (p+Pb\rightarrow p+Pb+V) & = & \frac{dN_{\gamma}^{Pb}(-y)}{d\omega }\sigma_{\gamma p \rightarrow V +p}(-y) \nonumber \\
& + &\frac{dN_{\gamma}^p(y)}{d\omega }\sigma_{\gamma Pb \rightarrow V +Pb}(y),\nonumber
\end{eqnarray}
where $\frac{dN_{\gamma}(y)}{d\omega }$ is the corresponding photon flux and $y = \ln (2\omega/M_V)$. The case for the inverse beam direction is straightforward. Similarly, the rapidity distribution $y$ in nucleus-nucleus collisions has the same factorized form,
\begin{eqnarray}
\frac{d\sigma}{dy} (A A \rightarrow   A\otimes V \otimes Y) & = &  \left[ \omega \frac{dN_{\gamma}^A}{d\omega }\,\sigma(\gamma A \rightarrow V +Y ) \right. \nonumber \\
& + & \left. \left(y\rightarrow -y \right) \right],
\label{dsigdyA}
\end{eqnarray}
where the photon flux in nucleus is denoted by $dN_{\gamma}^A/d\omega$ and $Y=A$ (coherent case) or $Y=A^*$ (incoherent case). The symbol $\otimes$ represents a large rapidity gap interval. 

\begin{figure*}[t]
\includegraphics[scale=0.45]{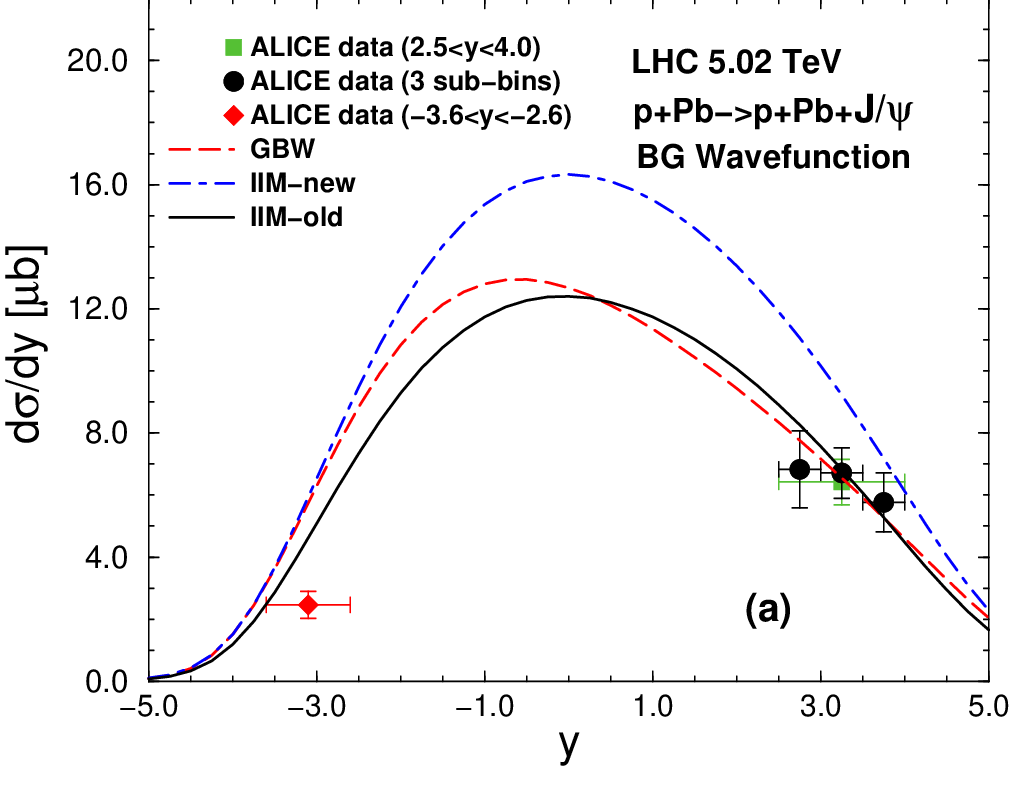} 
\includegraphics[scale=0.45]{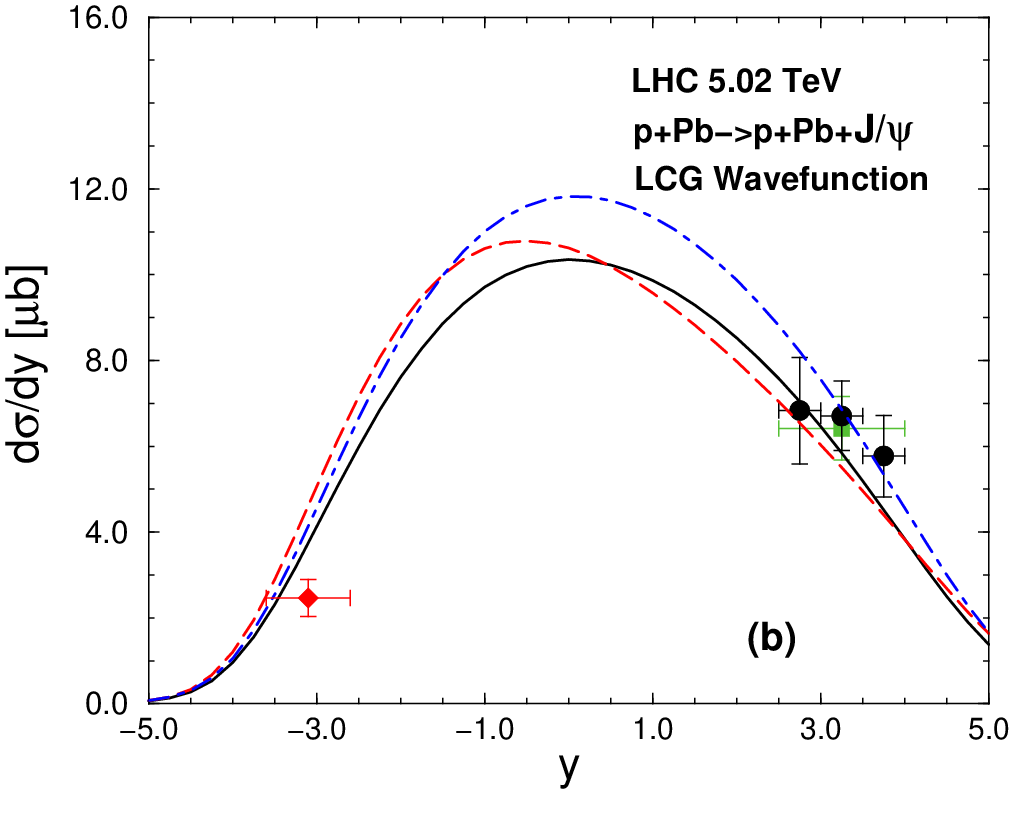}
 \caption{(Color online) Predictions for the rapidity distribution of $J/\psi$ photoproduction in $pPb$ collisions at LHC ($\sqrt{s_{NN}} = 5.02$ TeV) for the case of (a) Boosted Gaussian (BG) and (b) Light-Cone Gaussian (LCG) wavefunctions and several models for the dipole cross section (see text). The ALICE data \cite{ALICE3} for forward/backward rapidities are also included.}
\label{fig:1}
\end{figure*}

Here, the photon-Pomeron interaction will be described within the light-cone dipole frame, where the probing projectile fluctuates into a quark-antiquark pair with transverse separation $r$ (and momentum fraction $z$) long after the interaction, which then scatters off the hadron. The cross section for exclusive photoproduction of vector meson  off a nucleon target is given by \cite{Nemchik:1996pp} ,
\begin{eqnarray}
\label{sigmatot}
\sigma (\gamma p\rightarrow V p)  =   \frac{1}{16\pi B_V} \left|\int dz\, d^2r \,\Phi^{\gamma^*V}_T\sigma_{dip}(x,r)  \right|^2,
\end{eqnarray}
with
\begin{eqnarray}
\Phi^{\gamma^*V}_T  = \sum_{h, \bar{h}}\Psi^\gamma_{h, \bar{h}}(z,r,m_q)\Psi^{V*}_{h, \bar{h}}(z,r,m_q),
\end{eqnarray}
where $\Psi^{\gamma}(z,r,m_q)$ and $\Psi^{V}(z,r,m_q)$ are the light-cone wavefunction of the photon  and of the  vector meson, respectively.  The Bjorken variable is denoted by $x=M_V^2/(W_{\gamma p}^2-m_p^2)$, the dipole cross section by  $\sigma_{dip}(x,r)$ and the  diffractive slope parameter by $B_V$.  The centre-of-mass energy of photon-proton system is labeled by $W_{\gamma p}$. Here, we consider the energy dependence of the slope parameter using the Regge motivated expression, $B_V(W_{\gamma p}) = B_0+4\alpha^{\prime}\log(W_{\gamma p}/W_0)$. The values of parameters used are the following: for $J/\psi$ one has  $B_0=4.99$ GeV$^{-2}$, $\alpha^{\prime} = 0.25$ GeV$^{-2}$ and $W_0=90$ GeV and for $\Upsilon (1S)$ one has $B_0=3.68$ GeV$^{-2}$, $\alpha^{\prime} = 0.164$ GeV$^{-2}$ and $W_0=95$ GeV.

\begin{figure*}[t]
\includegraphics[scale=0.45]{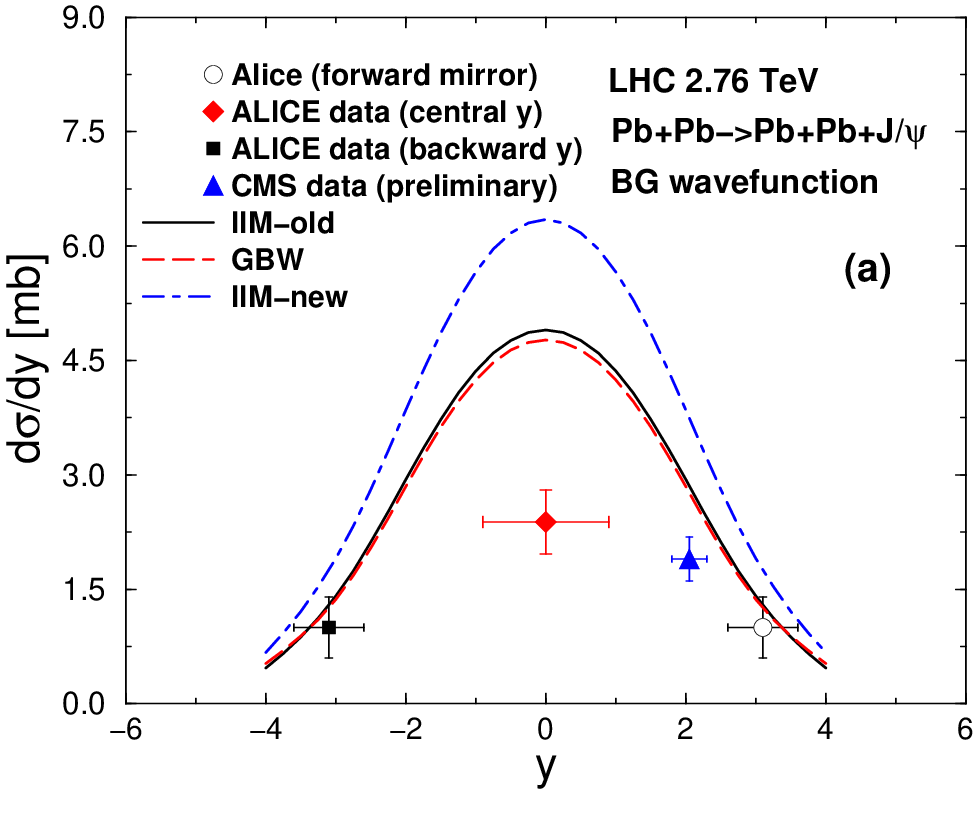} 
\includegraphics[scale=0.45]{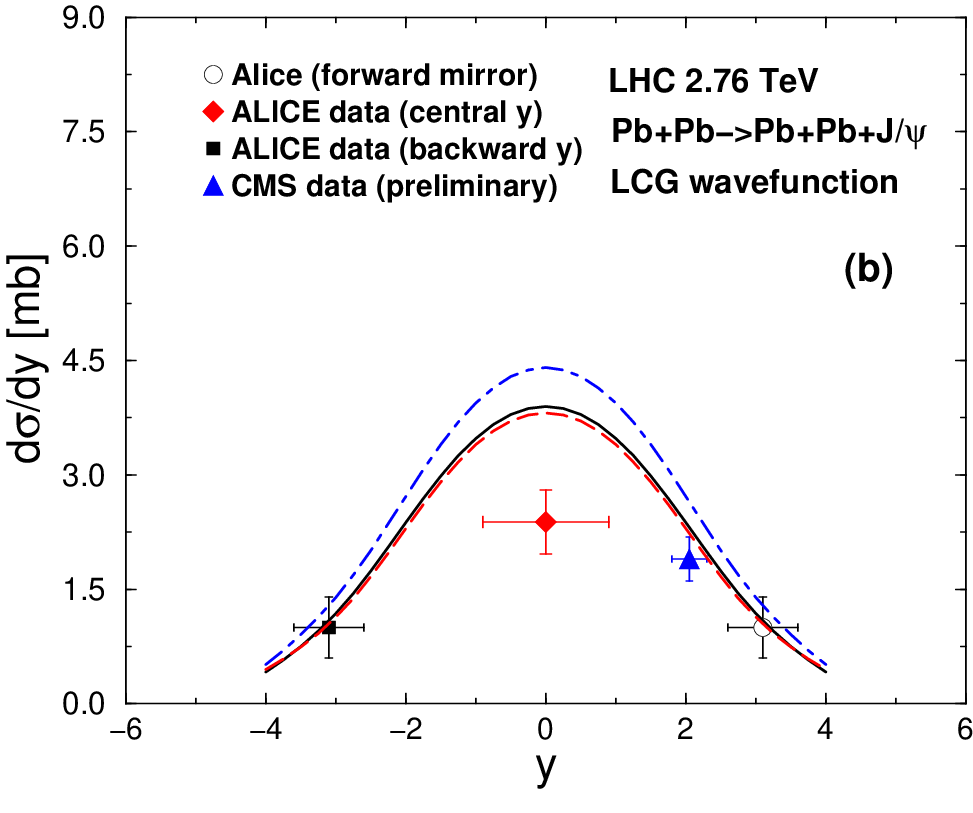}
 \caption{(Color online) Predictions for the rapidity distribution of $J/\psi$ photoproduction in $PbPb$ collisions at LHC ($\sqrt{s_{NN}} = 2.76$ TeV) for the case of (a) Boosted Gaussian (BG) and (b) Light-Cone Gaussian (LCG) wavefunctions and several models for the dipole cross section (see text). The ALICE data  for central \cite{ALICE1} and backward \cite{ALICE2}  rapidities are included. The preliminary CMS data \cite{CMSupc} is also shown.}
\label{fig:1-R}
\end{figure*}

Accordingly, the exclusive photoproduction off nuclei for coherent and incoherent processes can be simply computed in high energies where the large coherence length $l_c\gg R_A$ hypothesis is fairly valid. The expressions for both cases are given by \cite{Boris},
\begin{eqnarray}
\sigma (\gamma A \rightarrow VA ) & = & \int d^2b\, \left|\langle \Psi^V|1-\exp\left[-\frac{1}{2}\sigma_{dip} T_A \right]|\Psi^{\gamma}\rangle \right|^2, \label{eq:coher} \nonumber \\
\sigma (\gamma A \rightarrow VA^* )  & = & \frac{1}{16\pi\,B_V}\int d^2b\,T_A \left|\langle \Psi^V|\sigma_{dip}(x,r) \right. \nonumber \\
 &\times &  \left. \exp\left[-\frac{1}{2}\sigma_{dip}T_A(b)  \right]|\Psi^{\gamma}\rangle\right|^2, \nonumber
\label{eq:incoh}
\end{eqnarray} 
where $T_A(b)= \int dz\rho_A(b,z)$  is the nuclear thickness function. The notation $\langle \Psi^V| (\ldots)|\Psi^{\gamma}\rangle$ represents the overlap over the wavefunctions.

\section{Phenomenological inputs and models}

On source of theoretical uncertainty in such predictions is related to the choice for the meson wavefunction. In order to investigate this, we considered the Boosted Gaussian \cite{wfbg} (BG) and the Light-Cone Gaussian \cite{wflcg} (LCG). The expressions for
the overlap functions we have used appropriately summed over the helicity and flavor indices are given by:
\begin{eqnarray*}
\Phi^{\gamma^*V}_T(z,r,m_q)  & = &  \hat{e}_f \frac{\sqrt{4\pi\,\alpha_e}}{(2\pi)^2} N_c\left\{
             m_f^2 K_0(r\epsilon_f)\phi_T(r,z) \right. \nonumber \\
           & - & \left. [z^2+(1-z)^2]\,\epsilon_f K_1(r\epsilon_f) \partial_r\phi_T(r,z)
            \right\},
\end{eqnarray*}
where the constant $\hat{e}_f$ stands for an effective charge arising from the sum over quark flavors in the meson. It is
given in Table \ref{tab:wfparams} along with the quark and meson
masses used. Here,  $m_f$ denotes the mass of the quark with flavor $f$ 
and the quantity $\epsilon_f$ is given by $\epsilon_f = \sqrt{z(1-z)Q^2+m_f^2}$ (in photoproduction case, where the photon virtuality $Q^2 = 0$, $\epsilon_f = m_f$).  For the BG wavefunctions \cite{wfbg}, the function $\phi_T$ is given by,
\begin{eqnarray}
\phi_{T} = N_{T}\,\exp\left[-\frac{m_f^2R^2}{8z(1-z)}+\frac{m_f^2R^2}{2}-\frac{2z(1-z)r^2}{R^2}
\right]. \nonumber \\
\end{eqnarray}
The parameters $R$ and $N_{T}$ are constrained by unitarity of the
wavefuntion as well as by the electronic decay widths. They are given
in Table \ref{tab:wfparams}. On the other hand, for LCG wavefunction
\cite{wflcg} one has the following expression:
\begin{eqnarray}
\phi_T  =  N_T\,z(1-z)\,\exp\left[-r^2/(2R_T^2)\right],
\end{eqnarray}
with the parameters also given in Table \ref{tab:wfparams}. The parameters for the meson wavefunctions shown in Table I correspond to fixed heavy quark masses of $m_c=1.4$ GeV and $m_b=4.2$ GeV. See Refs. \cite{Sanda1,Sanda2,Kow1,Kow2,AmirArmesto} for similar determinations of wavefunction parameters for $J/\psi$ and $\Upsilon$ states. 

\begin{table}[t]
\begin{center}
\begin{tabular}{|c|c|c|c|c|c|c|c|}
\hline
             & \multicolumn{3}{|c|}{common parameters}
             & \multicolumn{2}{|c|}{BG parameters}
             & \multicolumn{2}{|c|}{LCG parameters}\\
\hline
$V$  & $M_V$ & $m_f$ & $\hat{e}_f$ &  $R^2$  & $N_T$ 
             & $R_T^2$  & $N_T$ \\
\hline
$J/\psi$  &   3.097 &  1.4 (1.27) & 2/$3$ & 2.44 & 0.572 & 6.5 & 1.23  \\
$\Upsilon (1S)$  &   9.46 &  4.2 (4.2) & 1/3 & 0.567  & 0.481  & 1.91 & 0.78  \\
\hline
\end{tabular}
\end{center}
\caption{Parameters for the vector-meson light-cone wavefunctions in units of GeV.  The numbers in parenthesis in the column $m_f$ are the heavy quark masses values considered in the IIM-new model \cite{Amir} (see discussion in text). }
\label{tab:wfparams}
\end{table}

Another source of theoretical uncertainty is the model for the dipole-target cross section. We used the parameterization proposed by Iancu, Itakura and Munier (IIM) \cite{IIM} (including charm quark in fits). In this case, the dipole cross section  is parameterized as follows,
\begin{eqnarray}
\sigma_{dip}\,(x,r) =\sigma_0\,\left\{ \begin{array}{ll} 
0.7 \left(\frac{\bar{\tau}^2}{4}\right)^{\gamma_{\mathrm{eff}}\,(x,\,r)}\,, & \mbox{for $\bar{\tau} \le 2$}\,,  \nonumber \\
 1 - \exp \left[ -a\,\ln^2\,(b\,\bar{\tau}) \right]\,,  & \mbox{for $\bar{\tau}  > 2$}\,, 
\end{array} \right.
\label{CGCfit}
\end{eqnarray}
where $\bar{\tau}=r Q_{\mathrm{sat}}(x)$. For the color transparency region near saturation border ($\bar{\tau} \le 2$), the behavior is driven by the effective anomalous dimension $\gamma_{\mathrm{eff}}\, (x,\,r)= \gamma_{\mathrm{sat}} + \frac{\ln (2/\tilde{\tau})}{\kappa \,\lambda \,y}$ with $\kappa = 9.9$. The saturation scale is defined as $Q_{sat}^2(x)=\left(\frac{x_0}{x}\right)^{\lambda}$ and $\sigma_0=2\pi R_p^2$.

The first set (labeled by IIM-old \cite{Soyez}) considers the previous DESY-HERA data and the values for parameters are $\gamma_{\mathrm{sat}}=0.7376$, $\lambda = 0.2197$, $x_0=0.1632\times 10^{-4}$ and $R_p= 3.344$ GeV$^{-1}$ ($\sigma_0=27.33$ mb). For IIM-old, the charm quark mass is fixed as $m_{c}=1.4$ GeV. The second set (labeled IIM-new \cite{Amir}) considered the extremely small error bars on the recent ZEUS and H1 combined results for inclusive DIS. In this case, the parameters are $\gamma_{\mathrm{sat}}=0.762$, $\lambda = 0.2319$, $x_0=0.6266\times 10^{-4}$ and $\sigma_0=21.85$ mb. For IIM-new, the charm quark mass is fixed as $m_{c}=1.27$ GeV. In order to compare the dependence on distinct models, we also consider the simple GBW parameterization \cite{GBW} (fit including charm quark). Notice that we have used the central values of fitted parameters. Their typical uncertainties are smaller than 0.5 percent and they do not produce substantial deviations on the presented predictions.

\begin{figure*}[t]
\includegraphics[scale=0.45]{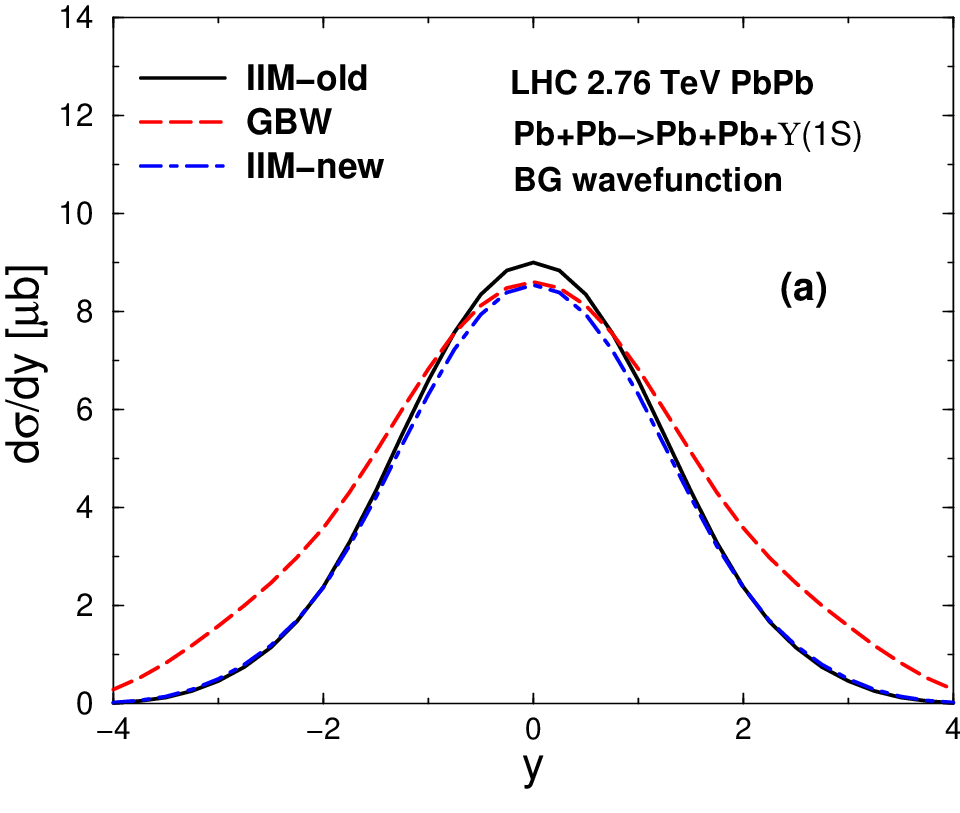} 
\includegraphics[scale=0.45]{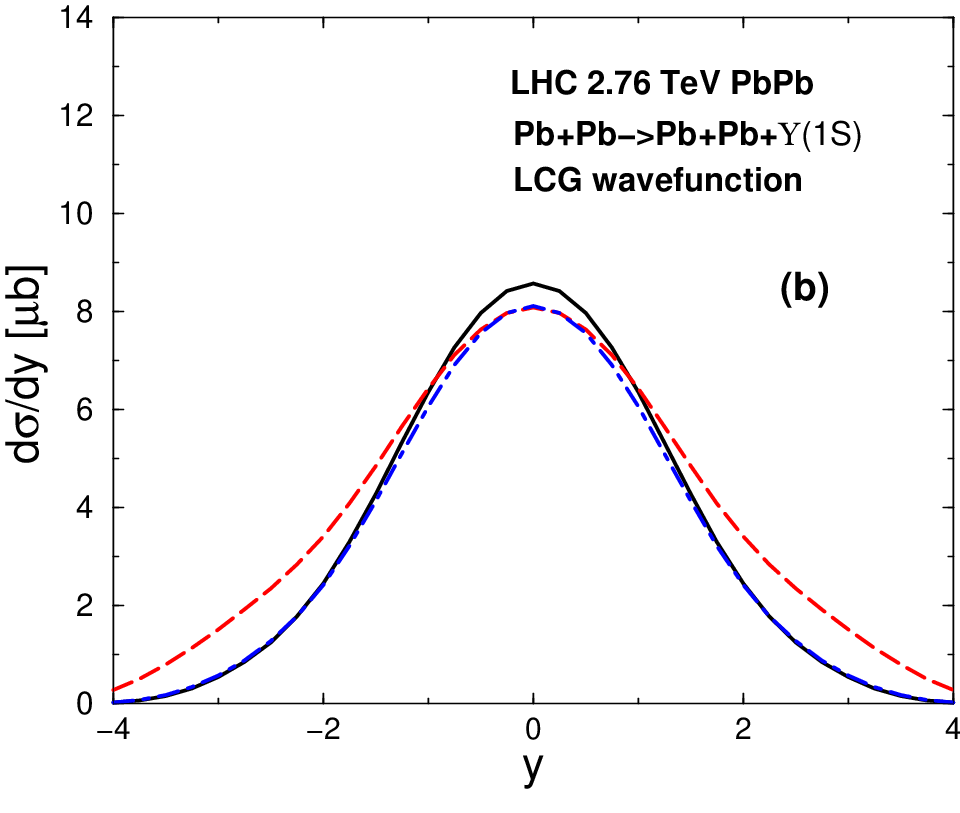}
\caption{(Color online) Predictions for the rapidity distribution of coherent photonuclear production of $\Upsilon(1S)$ in PbPb collisions at LHC ($\sqrt{s_{NN}} = 2.76$ TeV) for the case of (a)  Boosted Gaussian and (b) Light-Cone Gaussian (thin curves) wavefunctions and several models for the dipole cross section (see text).}
\label{fig:2}
\end{figure*}

As a final note on the details of the present calculation, we discuss the threshold correction, the real part of amplitude and skewness effects. In all numerical calculations, we multiply the dipole cross sections above by a threshold correction factor $(1-x)^n$, where $n = 7$ for the heavy ones (the value for $n$ is estimated using quark counting rules). The cross section in Eq. (\ref{sigmatot}) has been computed including the real part of amplitude contribution and skwedness correction in the following way, 
\begin{equation}
 \hat{\sigma}_{\gamma p\rightarrow Vp}=R_g^{2} \,{\sigma}(\gamma p\rightarrow Vp)(1+\beta^2) ,
  \label{vm}
\end{equation}
with
\begin{eqnarray}
 \label{eq:beta} 
  \beta & = &  \tan\left(\frac{\pi\varepsilon }{2}\right), \hspace{0.6cm}R_g(\varepsilon) = \frac{2^{2\varepsilon+3}}{\sqrt{\pi}}\frac{\Gamma(\varepsilon+5/2)}{\Gamma(\varepsilon+4)}, \nonumber\\
\varepsilon &\equiv& \frac{\partial\ln\left(\mathcal{A}^{\gamma p\rightarrow Vp}\right)}{\partial\ln(1/x)}, \
\end{eqnarray}
where the factor $(1+\beta^2)$ takes into account the missing real part of amplitude, with $\beta$ being the ratio of real to imaginary parts of the scattering amplitude.  The factor $R_g$ incorporates the skewness effect, coming from the fact that the gluons attached to the $q\bar{q}$ can carry different light-front fractions $x,x^{\prime}$ of the proton.
The skewedness factor given in Eq. (\ref{eq:beta})  was obtained at NLO level, in the limit that $x^{\prime}\ll x\ll 1$ and at small $t$ assuming that the diagonal gluon density of target has a power-law form  \cite{skew}. Similar corrections have been introduced also for the photonuclear cross section.

As a comment on the reliability of the present calculation, the non-forward wavefunction \cite{Bartels} was not considered as the dipole cross sections used have not explicit impact-parameter or $t$-dependence. There are few phenomenological models in literature considering b-dependence (IP-saturation model \cite{IP} and CGC impact parameter model (b-CGC) \cite{bCGC}) and they basically take into account Gaussian-like behavior associated to the proton wavefunction (this is enough to describe the small-$t$ region from data on $ep$ exclusive processes). The same comment applies to the Marquet-Peschanski-Soyez (MPS) model \cite{MPS}, where explicit $t$-dependence for dipole cross section is considered. In any case, the IIM  dipole cross section considered in our calculation is a limit case ($t \rightarrow 0$) for both  CGC impact parameter and MPS models. The more complex question of b-dependence coming from the evolution equations for the dipole amplitude still remains an open question (see, e.g.  Ref. \cite{Berger}). We considered a simpler approach, where the $B_V$ slope is taken from a parameterization  for the measured values at HERA and we consider only the forward dipole amplitude (this diminishes the theoretical uncertainty associated to the specific model for the b-dependence in dipole amplitude). It would be opportune to compare our calculation to the results from Ref. \cite{AmirArmesto}, where the photoproduction of $J/\psi$ has been considered using both the IP-saturation and the b-CGC models. Concerning the expression for the dipole-nucleus cross section, we are using the Glauber-Gribov (smooth nucleus) approach which has been tested against small-$x$ data for nuclear structure functions (see, e.g. Ref.  \cite{Armesto}). A more refined treatment of nuclear effects has been done in Ref. \cite{IP}, where the gluons form a lumpy distribution within the nucleus. To account for this correlation among the gluons those authors generalize the smooth nucleus model. It was shown that only if the atomic mass  number $A$ is large and dipole size $r$ is small  the smooth nucleus formula is recovered. This is the case for the calculation we are considering: lead nucleus and heavy meson production with typical dipole sizes given by $r \sim 1/(Q^2+m_V^2)$. 

\section{Main results}

\begin{figure*}[t]
\includegraphics[scale=0.45]{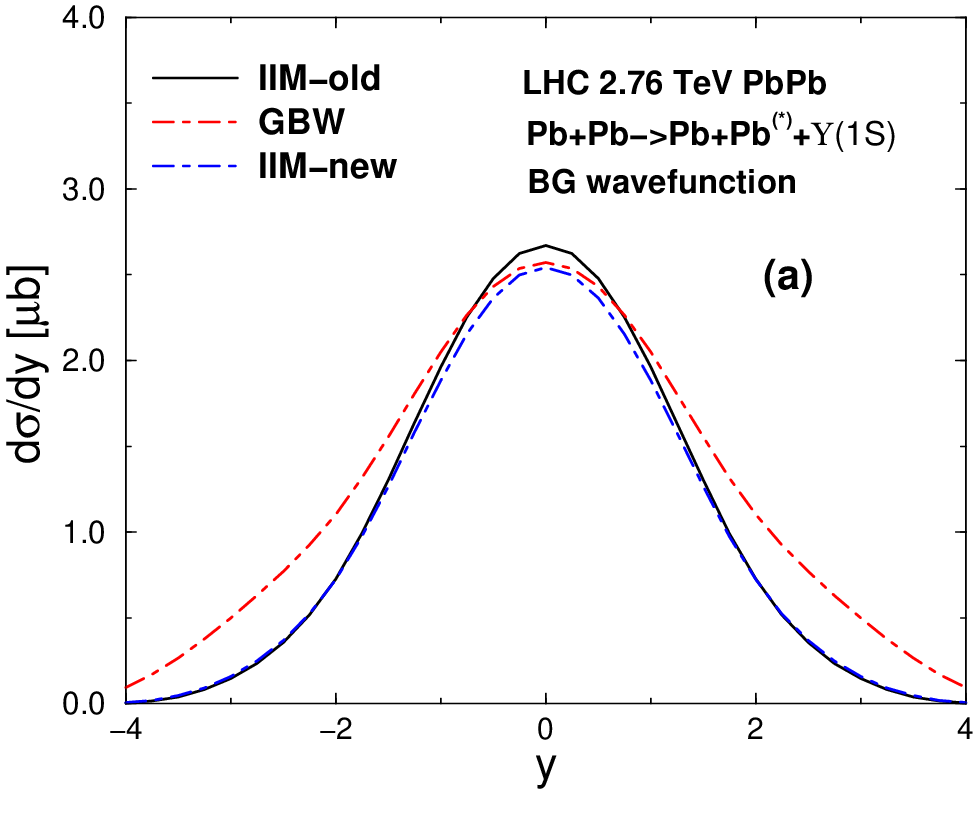} 
\includegraphics[scale=0.45]{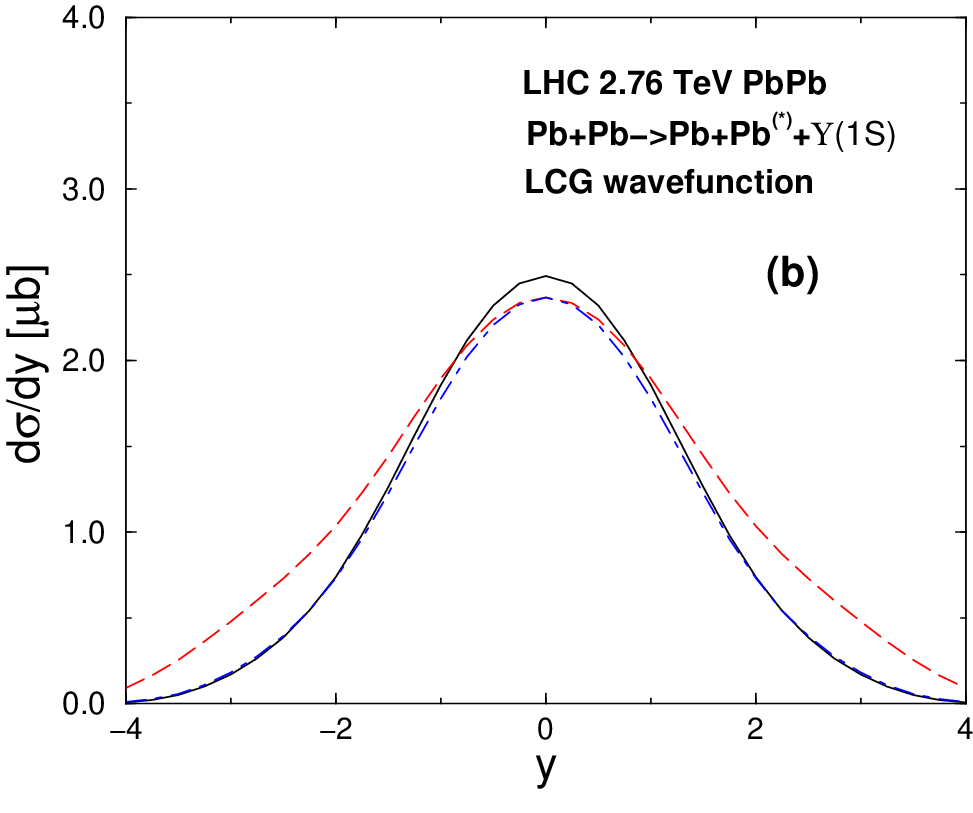} 
 \caption{(Color online) Predictions for the rapidity distribution of incoherent photonuclear production of $\Upsilon(1S)$ in PbPb collisions at LHC ($\sqrt{s_{NN}} = 2.76$ TeV) for the case of  (a) Boosted Gaussian  and (b) Light-Cone Gaussian wavefunctions and several models for the dipole cross section (see text).}
\label{fig:3}
\end{figure*}
 
We are now ready to obtain the rapidity distribution for $J/\psi$ photoproduction in proton-lead collisions at $\sqrt{s_{NN}}= 5.02$ TeV at the LHC. In Fig. \ref{fig:1}-(a)  is presented the results for $J/\psi$ taking into account the BG wavefunction and some samples of phenomenological models for the dipole cross section as discussed before. The solid curve stands for the IIM-old dipole cross section, whereas the dotted-dashed line represents the result using the new fit, IIM-new. The dashed curve stands for the GBW parametrisation. The behavior at very large rapidities tends to be similar for the distinct models. However, towards midi-rapidity there is an evident model dependence. The IIM-old and GBW parameterization deliver quite similar results, whereas IIM-new produces a overall normalization large than the others. This is mainly due the smaller values of the charm quark mass considered ($m_c=1.27$ GeV in contrast to $m_c=1.4$ GeV for IIM-old and GBW). Also shown are the ALICE data \cite{ALICE3} for forward rapidities: the wide range $2.5<y<4.0$ (filled square) and three sub-bins y:[3.5,4.0],[3.0,3.5] and [2.5,3.0] (filled circles). The statistical and systematic uncertainties have been summed in quadrature.  In Fig. \ref{fig:1}-(b), the results are now presented for the LCG wavefunction and the notation is the same as for the previous plot. When using the LCG, the IIM-new parameterization provides a better description of the experimental data. Also, a smaller deviation between IIM-old and IIM-new parameterizations is now observed. This is due the difference in the $\phi_T$ function, which in the case of the LCG wave function does not depend explicitly on $m_c$. The predictions using the LCG wavefunction are in general smaller than the BG wavefunction case. The ALICE collaboration has also measured the rapidity cross section for the case $p+Pb\rightarrow p+Pb+J/\psi$ in the region $-3.6<y<-2.6$, where $d\sigma/dy (-3.6<y<-2.6) = 2.46\pm 0.31\, (\mathrm{stat}) \,^{+0.23}_{-0.27}\,(\mathrm{syst})$  $\mu$b. We include it in Fig. \ref{fig:1} and once again the LCG wavefunction is in better agreement with data compared to the BG wafefunction.  As a final remark on this point, our results are somewhat consistent with those  presented in Ref. \cite{Lapi} where the $Pb+p\rightarrow Pb+p+J/\psi$ cross section was computed using a distinct treatment for the coherent and incoherent amplitudes. 

For sake of comparison, in Fig. \ref{fig:1-R} we present the results for the the rapidity distribution of $J/\psi$ photoproduction in $Pb+Pb$ collisions at LHC   for $\sqrt{s_{NN}} = 2.76$ TeV. We consider the case of BG (Fig. \ref{fig:1-R}-a) and the LCG wavefunctions (Fig. \ref{fig:1-R}-b). The labels are the same as for Fig. \ref{fig:1}.  The ALICE data  for central \cite{ALICE1} and backward \cite{ALICE2}  rapidities are included (we take into account the mirror of this data point in forward rapidities for sake of visualization). The preliminary CMS data \cite{CMSupc} in forward rapidities is also shown.  The conclusions are the same as for $p+Pb$ collisions: the uncertainty related to the meson wavefunction is large, mostly at central rapidities. The LCG wavefuntion is preferred in comparison to the BG wavefunction (despite the model for dipole cross section).

\begin{figure*}[t]
\includegraphics[scale=0.45]{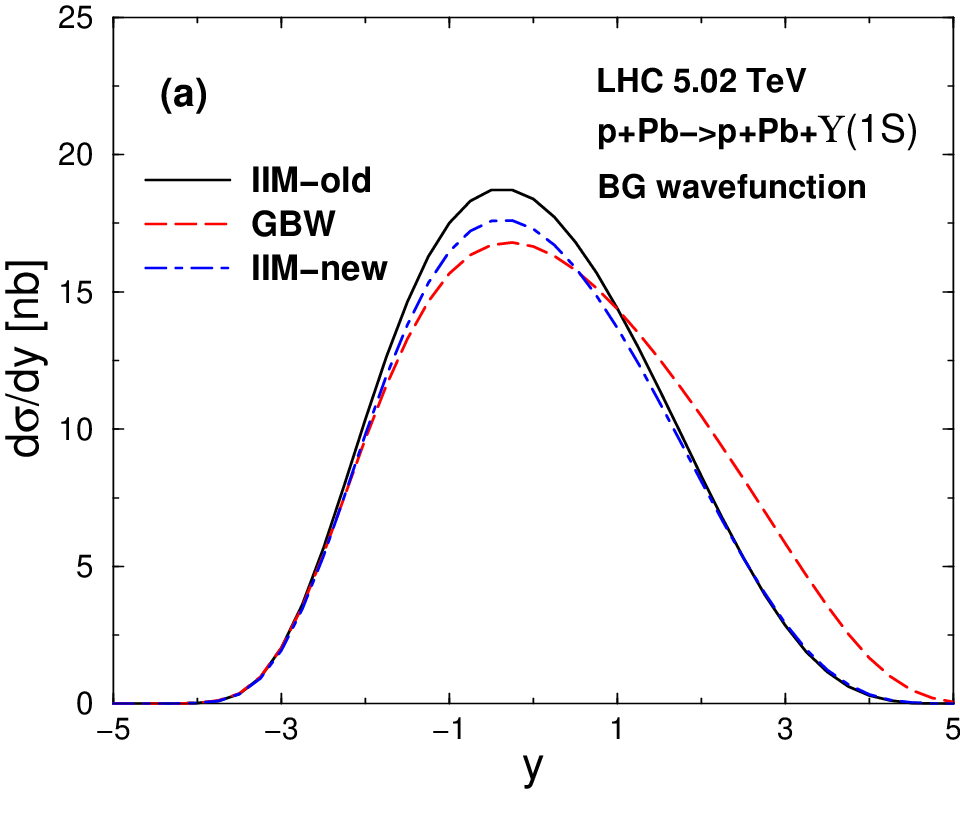} 
\includegraphics[scale=0.45]{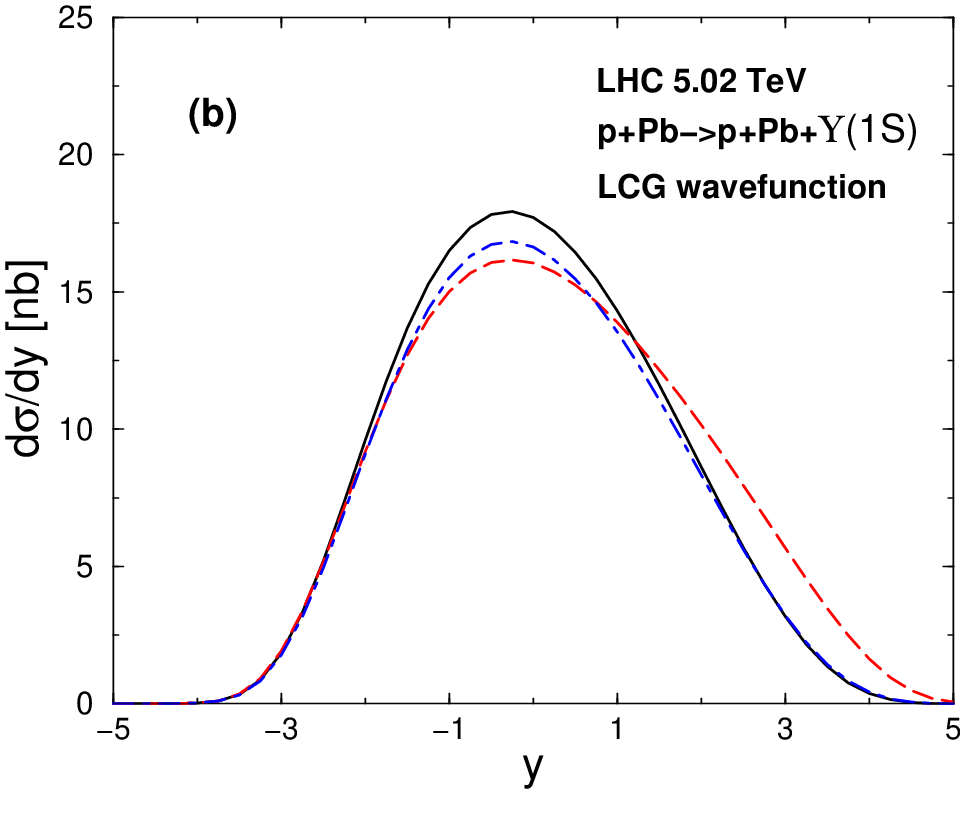} 
 \caption{(Color online) Predictions for the rapidity distribution for  production of $\Upsilon(1S)$ in pPb collisions at LHC ($\sqrt{s_{NN}} = 5.02$ TeV) for the case of  (a) Boosted Gaussian  and (b) Light-Cone Gaussian wavefunctions and several models for the dipole cross section (see text).}
\label{fig:3-R}
\end{figure*}

\begin{figure}[t]
\includegraphics[scale=0.45]{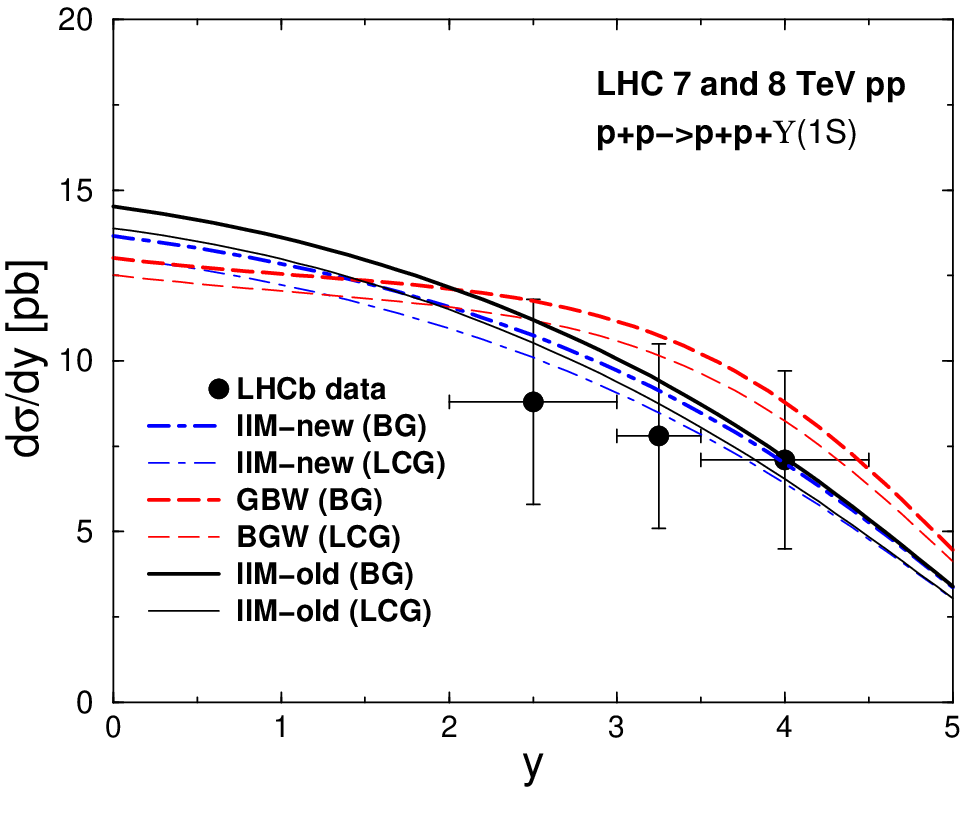} 
 \caption{(Color online) Predictions for the rapidity distribution for  production of $\Upsilon(1S)$ in proton-proton collisions at LHC ($\sqrt{s} = 7$ and 8 TeV) for the case of  Boosted Gaussian (bold curves)  and Light-Cone Gaussian (thin curves) wavefunctions and several models for the dipole cross section. Data  from LHCb Collaboration \cite{UpsilonLHCb} is also presented. }
\label{fig:4}
\end{figure}

Let us now investigate the photonuclear production of $\Upsilon$  in nucleus-nucleus collisions at the LHC. We will consider PbPb  collisions at the energy of 2.76 TeV. In Fig. \ref{fig:2} we present the results for the rapidity distributions for the coherent $\Upsilon$ production, $\mathrm{Pb}+\mathrm{Pb}\rightarrow \mathrm{Pb}+\Upsilon (1S)+\mathrm{Pb}$, considering both wavefunctions (without nuclear break up). Curves in Fig. \ref{fig:2}-a correspond to the result using BG wavefunction and the predictions using LCG wavefunction are shown in Fig. \ref{fig:2}-b. The theoretical uncertainty associated to the model for wavefunction is different compared to the pA case and now the results are similar. The main reason is that we are considering the same mass for the bottom quark , $m_b=4.2$ in both parameterizations. It is verified that the deviation is significant concerning the distinct models for the dipole cross section mostly at large rapidities.  In Fig. \ref{fig:3} one presents the results for the incoherent production of $\Upsilon$, $\mathrm{Pb}+\mathrm{Pb}\rightarrow \mathrm{Pb}+\Upsilon (1S)+\mathrm{Pb}^*$ taking into account the BG (Fig. \ref{fig:3}-a)  and LCG (Fig. \ref{fig:3}-b) wavefunctions. The notation in Fig. \ref{fig:3}  is the same as for Fig. \ref{fig:2}. As for the coherent case, the deviation is quite small when comparing the dependence on the meson wavefunction.  We note that we have not introduced the effect of nuclear break up  as we are considering here events with no neutron emitted in any direction ($0n0n$). The strong fields associated with heavy-ions lead to large probabilities for exchanging additional photons when a $\Upsilon$ meson is produced at small impact parameters and they will excite one or both of the nuclei and lead to break up.  For sake of comparison, in Fig. \ref{fig:3-R} we present the results for the the rapidity distribution of $\Upsilon (1S)$ photoproduction in $p+Pb$ collisions at LHC   for $\sqrt{s_{NN}} = 5.02$ TeV. We consider the case of BG (Fig. \ref{fig:3-R}-a) and the LCG wavefunctions (Fig. \ref{fig:3-R}-b). The labels are the same as for Figs. \ref{fig:2} and \ref{fig:3}. The theoretical uncertainties concerning the models for meson wavefunction and dipole cross sections are less sizable when compared to the $J/\psi$ case. Here, we consider only the contribution from $\sigma (\gamma p\rightarrow \Upsilon + p)$ (the photonuclear contribution is disregarded) and the slope parameter  is given by $B_{\Upsilon}(W_{\gamma p}) = 3.68+4(0.164)\log(W_{\gamma p}/95 \,\mathrm{GeV})$. In Ref. \cite{Glauber1}, the cross section $\sigma (\gamma p\rightarrow \Upsilon + p)$ has been multiplied by a factor 2.16 in order to describe the correct normalization of the old DESY-HERA data. For sake of completeness we also present the corresponding results for the proton-proton case in Fig. \ref{fig:4} (rapidity distribution for forward $y$) and compare  them to the recently available data from LHCb collaboration \cite{UpsilonLHCb}. The thin curves are for the LCG wavefunction and bold curves are for BG wavefunction. The uncertainty regarding the choice of wavefunction seems similar to the one for nucleus-nucleus and proton-nucleus cases.

Finally, we discuss the possible limitations of current approach. Here, we ignore the important problem with simultaneous description of the ALICE data on $J/\psi$ photoproduction in coherent and incoherent Pb-Pb Ultra-Peripheral Collisions (UPC) in the color dipole approach. It was shown in Ref. \cite{GGM2} that the color dipole model dramatically overestimates the $y=0$ ALICE data point on coherent $J/\psi$ photoproduction in Pb-Pb UPCs at 2.76 TeV and that the agreement can be achieved only by introducing an {\it ad hoc} gluon shadowing factor of $R_G$. At the same time, the $y=0$ ALICE data point on incoherent $J/\psi $ photoproduction in Pb-Pb UPCs at 2.76 TeV is reproduced by the dipole formalism assuming $R_G=1$. However, in Ref. \cite{GGM2} only the BG wavefunction and IIM-old dipole model have been considered. As discussed before, when the LCG wavefunction is considered the results are smaller than the BG case. Thus, there is some space to account for the discrepancy between data and dipole approach results in the theoretical uncertainties associated to the choices for the wavefunctions, the model for the dipole cross section and the values of heavy quark masses. In fact, we have shown in \cite{Glauber1} that for $pA$ collisions the effect of an additional $R_G$ contribution is negligible for practical situations. In addition, while in $pp$ and $AA$ UPCs, the photon flux is known rather well, the case of $pA$ UPCs is less clear since the photon flux at large photon energies is significantly suppressed by soft $pA$ interactions at
small impact parameters as discussed in Ref. \cite{Guzey}. We have not considered this uncertainty in the present calculation.

\section{Summary}

Summarizing, an investigation was done on the  photoproduction of $J/\psi$  and $\Upsilon (1S)$ meson states in the proton-nucleus and nucleus-nucleus collisions at LHC energies. It was included both contributions of photon-proton and proton-nucleus interactions within the color dipole formalism. Predictions for the rapidity distributions are presented  using the color dipole formalism and including saturation effects that are expected to be relevant at high energies.  Predictions for the rapidity distributions are presented and the dependence on the meson wavefunction, heavy quark mass as well as the models for the dipole cross section are analyzed. We compare directly the theoretical results to the recent data from ALICE collaboration on $J/\psi$ photoproduction in $p+Pb$ collisions at $\sqrt{s_{NN}} = 5.02$ TeV and in $Pb+Pb$ collisions at $\sqrt{s_{NN}} = 2.76$ TeV. Predictions are also performed for $\Upsilon (1S)$ state in $Pb+Pb$ and $p+Pb$ collisions at the LHC energies, including the coherent and incoherent contributions. The uncertainties concerning the model for the meson wavefunction is very large, despite the model considered for the dipole cross section. This is more significant for $Pb+Pb$ and $p+Pb$ collisions in the $J/\psi$ case, mostly at central rapidities. The experimental results select the LCG wavefunction as the better option. Concerning the dependence on the dipole cross section, the IIM-old and IIM-new parameterizations are not directly comparable as their parameters are fitted using distinct values for the charm quark mass. This introduces an additional non-trivial cross-dependence on $m_c$ in the wavefunctions (BG depends explicitly on the heavy-quark masses). The situation is in much better control for $\Upsilon$ case, where the considered bottom mass is the same.

\begin{acknowledgments}
This work was  partially financed by the Brazilian funding agency CNPq and by the French-Brazilian scientific cooperation project CAPES-COFECUB 744/12. MVTM thanks Amir Rezaeian for useful discussions.
\end{acknowledgments}

\end{document}